\documentclass[10pts]{article}
\usepackage{amsthm}
\theoremstyle{plain}
\usepackage{graphicx} 
\usepackage{braket}
\newcommand{\kernel}[1]{\left<q\left.\left|#1\right.\right|q'\right>}
\newcommand{\opr}[1]{\mathsf{#1}}
\newcommand{\abs}[1]{\left|#1\right|}
\newtheorem*{theorem*}{Theorem}
\usepackage{amsfonts}
\begin{document}

\title{\bf The Bender-Dunne basis operators  as Hilbert space operators}
\author{Joseph Bunao and Eric A. Galapon\thanks{eagalapon@up.edu.ph, eric.galapon@upd.edu.ph}  \\ Theoretical Physics Group, National Institute of Physics\\University of the Philippines, 1101 Philippines}
\maketitle
\begin{abstract}
The Bender-Dunne basis operators, $\opr{T}_{-m,n}=2^{-n}\sum_{k=0}^n { n \choose k} \opr{q}^k \opr{p}^{-m} \opr{q}^{n-k}$ where $\opr{q}$ and $\opr{p}$ are the position and momentum operators respectively,  are formal integral operators in position representation in the entire real line $\mathbb{R}$ for positive integers $n$ and $m$. We show, by explicit construction of a dense domain, that the operators $\opr{T}_{-m,n}$'s are densely defined operators in the Hilbert space $L^2(\mathbb{R})$.
\\
\\
{\bf Keywords:} Time operators, Weyl quantization
\\
{\bf PACS:} 03.65.Xp, 02.30.Sa, 02.30.Uu
\end{abstract}

\section{Introduction}
In \cite{bender1,bender2} Bender and Dunne considered the problem of solving the Heisenberg equation of motion $i \hbar \dot{\opr{q}}= [\opr{q},\opr{H}]$, $i\hbar\dot{\opr{p}}=[\opr{p},\opr{H}]$ for the quantum mechanical Hamiltonian $\opr{H}=H(\opr{q},\opr{p})$ in one degree of freedom, where $\opr{q}$ and $\opr{p}$ are the position and momentum operators, respectively. They obtained an implicit solution to the problem by solving the operator equation $[F(\opr{q},\opr{p}),H(\opr{q},\opr{p})]=i\hbar\opr{I}$ for $F(\opr{q},\opr{p})$. Their solution constitutes expanding $H(\opr{q},\opr{p})$ and $F(\opr{q},\opr{p})$ in terms of some basis operators, $\opr{T}_{m,n}$, i.e. $H(\opr{q},\opr{p})=\sum_{m,n} h_{m,n} \opr{T}_{m,n}$ and $F(\opr{q},\opr{p})=\sum_{m,n}\alpha_{m,n}\opr{T}_{m,n}$, where the coefficients $h_{m,n}$ are known for a given Hamiltonian and the $\alpha_{m,n}$'s are to be determined. The basis operators are the Weyl-ordered quantizations of the one-dimensional monomials of the classical position and momentum observables, $q^n p^m$, for integers $n$ and $m$. Explicitly, when $m\geq 0$ and $n\geq 0$, the $\opr{T}_{m,n}$'s are either given by the following expressions
\begin{equation}\label{exp1}
\opr{T}_{m,n}=\frac{1}{2^n}\sum_{k=0}^n { n \choose k} \opr{q}^k \opr{p}^m \opr{q}^{n-k},
\end{equation}
\begin{equation}\label{exp2}
\opr{T}_{m,n}=\frac{1}{2^m} \sum_{j=0}^m { m \choose j} \opr{p}^j \opr{q}^n \opr{p}^{m-j} .
\end{equation}
These expressions can be extended to accomodate the case when either $m$ or $n$ is negative. When $n\geq 0$ and $m<0$ ($m\geq 0$ and $n<0$), the corresponding basis operator is given by equation \ref{exp1} (equation \ref{exp2}). 
However, Bender and Dunne recognized the formality of the progress they had reported because the operator solutions $F(\opr{q},\opr{p})$ could be ``extremely singular'' as they contained arbitrary inverse powers of $\opr{q}$ and $\opr{p}$ coming from the $\opr{T}_{m,n}$'s. With that they, too, recognized the need to clarify the actions of the operators $\opr{T}_{m,n}$ in the Hilbert space of a one-dimesional quantum particle or their status as Hilbert space operators. Doing so could lift the formality of the solutions $F(\opr{q},\opr{p})$, which could pave a way to solve spectral problems of Schrodinger operators. 

Meanwhile, in \cite{galapon1} Galapon, on offering a solution to the quantum time of arrival problem \cite{sombillo,freetoa,freetoa2,grot} in the interacting case \cite{oth8}, tackled the problem of constructing a time of arrival operator $\opr{T}$ for a given arrival point $x=0$ in the configuration space without the aid of quantization. The problem constitutes finding a solution to the Time-Energy canonical commutation relation $[\opr{T},\opr{H}]=i\hbar\opr{I}$ for a given Hamiltonian $\opr{H}$ under the condition that the solution $\opr{T}$ reduces, in the classical limit, to the classical time of arrival
\begin{equation}\label{ctoa}
T_{0}(q,p)=-\mbox{sgn}(p)\sqrt{\frac{\mu}{2}} \int_0^q \frac{dq'}{\sqrt{H(q,p)-V(q')}} .
 \end{equation}
That is $\opr{T}=T(\opr{q},\opr{p})$ reduces to $T_0(q,p)$ in the limit of commuting $\opr{q}$ and $\opr{p}$ or as $\hbar\rightarrow 0$.   Clearly the problem of constructing the time of arrival operator $\opr{T}$ is related to the problem earlier solved by Bender and Dunne, except that the solution $\opr{T}$ must satisfy the correspondence principle which the solution $F(\opr{q},\opr{p})$ does no have to satisfy. 
Moreover, the problem for $\opr{T}$ was sought in position representation so that it was not explicitly expressed in terms of position and momentum operators but expressed as an integral operator in position space. That is the time of arrival operator $\opr{T}$ is given by the integral operator
\begin{equation}\label{qtoa}
 \left(\opr{T}\varphi\right)\!\!(q)=\int_{-\infty}^{\infty}\kernel{\opr{T}} \varphi(q')\, \mbox{d}q',
\end{equation}
where $\kernel{T}=\mu T(q,q') \mbox{sgn}(q-q')/4 i \hbar$, in which $T(q,q')$ is the solution to the second order partial differential equation
\begin{equation}
-\frac{\hbar^2}{2 \mu} \frac{\partial^2 T(q,q')}{\partial q^2}+\frac{\hbar^2}{2 \mu} \frac{\partial^2 T(q,q')}{\partial q'^2} + \left(V(q)-V(q')\right) T(q,q')=0,
\end{equation}
subject to the boundary conditions $T(q,q)=q/2$ and $T(q,-q)=0$. 
For everywhere continuous potentials, $V(q)$, a unique solution exists. The boundary conditions guarantee that the time of arrival operator $\opr{T}$ reduces to the classical time of arrival in the classical limit. 

Despite the lack of obvious relationship between the time of arrival operator $\opr{T}$ to the Bender-Dunne basis operators $\opr{T}_{m,n}$, the $\opr{T}_{m,n}$'s are furtively embedded in $\opr{T}$. For everywhere continous potentials,  the kernel is given by $\kernel{\opr{T}}=\kernel{\opr{T}_W}+\kernel{\Delta \opr{T}}$, where $\opr{T}_W$ is the Weyl quantized version of the classical time of arrival and $\Delta \opr{T}$ is the quantum correction to $\opr{T}_W$. The operator $\opr{T}_W$ is obtained by expanding the classical time of arrival $T_0(q,p)$ in terms of the monomials $\opr{q}^n\opr{p}^{-m}$ for positive $m$ and $n$ , and then Weyl-quantizing the resulting series by affecting the replacement $q^n p^{-m}\mapsto \opr{T}_{-m,n}$. The transition to the kernel follows from the identity $\kernel{\opr{T}_{-m,n}}=2^{-n}\left(q+q'\right)^n \kernel{\opr{p}^{-m}}$,
where use have been made of the fact that $\ket{q}$ is an eigenvector of the position operator. The kernel $\kernel{\opr{p}^{-m}}$ is evaluated in momentum space, with the resulting integral interpreted as a distributional integral. The result is
\begin{equation}\label{kernelmn}
\kernel{\opr{T}_{-m,n}}= \frac{i(-1)^{(m-1)/2}}{2^{n+1}\hbar^{m}(m-1)!} (q+q')^{n}(q-q')^{m-1}\mbox{sgn}(q-q') .
\end{equation}
In position representation, the Bender-Dunne operators $\opr{T}_{-m,n}$ are then the integral operators,
\begin{equation}\label{bd}
\left(\opr{T}_{-m,n}\varphi\right)\!(x)=\int_{-\infty}^{\infty} \kernel{\opr{T}_{-m,n}}\varphi(q')\, \mbox{d}q'
\end{equation}
Moreover, the time of arrival operator $\opr{T}$ can be obtained in explicit operator by solving the operator equation $[\opr{T},\opr{H}]=i\hbar \opr{I}$ subject to the condition that $\opr{T}$ reduces to $T_0(q,p)$ in the limit of commuting $\opr{q}$ and $\opr{p}$. The solution $\opr{T}$ can be obtained by means of Bender and Dunne's method, expanding $\opr{T}$ in terms of the Bender-Dunne operators $\opr{T}_{m,n}$ and determining the expansion coefficients. The time of arrival operator in position representation is then obtained by using equation \ref{bd} \cite{galapon2}. 

Initial investigations on the properties of the time of arrival operator was done by spatial confinement, followed by successively increasing the confining length to discern  properties of the time of arrival operator in the entire configuration space \cite{galapon3,galapon4,galapon5}. This  allowed the construction of time arrival distributions and gave insights into the dynamics of the (confined) time of arrival operator eigenfunctions. The eigenfunctions unitarily evolve such that the eigenfunctions become most localized at the arrival point at the time corresponding to their eigenvalues, with the localization increasing with the confining length. This dynamical behavior of the eigenfunctions offers a novel theory of the appearance of particle as explored in \cite{galapon7}. More recently, the time of arrival operators have been applied in the entire real line for the free \cite{galapon8} and tunneling case \cite{galapon9}; in these works, expected times of arrivals have been computed as expectation values of the time of arrival operator. However, until the status of the time of arrival operator $\opr{T}$ is unclear as a Hilbert space operator in  $L^2(\mathbb{R})$, the full physical content of the theory of quantum arrival embodied in $\opr{T}$ remains to be unravelled.

As a first step in clarifying the Hilbert space status of the Bender-Dunne solutions $F(\opr{q},\opr{p})$ and the time of arrival operator $\opr{T}$, it is essential to establish that the formal Bender-Dunne operators $\opr{T}_{m,n}$ are meaningful operators in the Hilbert space $L^2(\mathbb{R})$. This requires demonstrating that the $\opr{T}_{m,n}$'s have non-trivial dense domains, i.e. there exists a dense subspace of $L^2(\mathbb{R})$ that is mapped by $\opr{T}_{m,n}$ into $L^2(\mathbb{R})$. For positive integers $m$ and $n$, the Bender-Dunne operators are clearly Hilbert space operators with the  Schwartz space $\mathcal{S}(\mathbb{R})\subset L^2(\mathbb{R})$ as their common domain. Here we limit ourselves to negative powers of the momentum operator and to positive powers of the position operator; this case is most relevant to the Bender-Dunne (minimal) solutions and to quantum time of arrival problem, and in the quantum time problem in general \cite{time1,time2}. It is then the objective of this paper to explictly establish in detail the following
\begin{theorem*}
For fixed positive integers $m$ and $n$ the operator $\opr{T}_{-m,n}$ is a densely defined operator in the Hilbert space $L^{2}(\mathbb{R})$. 
\end{theorem*}

The method of proof will consist of three parts. First is the indentification of necessary and sufficient conditions for a vector $\varphi(q)\in L^2(\mathbb{R})$ to be mapped by $\opr{T}_{-m,n}$ into $L^2(\mathbb{R})$. Second is the construction of a complete set of vectors of $L^2(\mathbb{R})$, which we denote by $S(\opr{T}_{-m,n})$, that are mapped by $\opr{T}_{-m,n}$ into the Hilbert space. The set $S(\opr{T}_{-m,n})$ in fact consists of a denumerable set of complete set of vectors.  We will refer to the elements of $S(\opr{T}_{-m,n})$ as basic vectors. Third is the assignment of the linear span of $S(\opr{T}_{-m,n})$ as the domain, $D(\opr{T}_{-m,n})$, of $\opr{T}_{-m,n}$, which is made possible by the linearity of $\opr{T}_{-m,n}$.  The proof now follows from the fact that, since $S(\opr{T}_{-m,n})$ is a complete set, the linear span,  $D(\opr{T}_{-m,n})$, of $S(\opr{T}_{-m,n})$ is necessarily dense in $L^2(\mathbb{R})$, so that the operator $\opr{T}_{-m,n}:D(\opr{T}_{-m,n})\subset L^2(\mathbb{R})\mapsto L^2(\mathbb{R})$ is necessarily a densely defined operator  in the Hilbert space $L^{2}(\mathbb{R})$.

Before we proceed, we mention that aside from the original applications of the operators $\opr{T}_{m,n}$'s due to Bender and Dunne and their later application to the quantum time of arrival problem, the Bender-Dunne operators have appeared in various applications. Closely related, the Bender-Dunne operators were needed to solve for the homogeneous solution to the time-energy commutation relation and it turns out that this solution need not be a function of the Hamiltonian (the energy operator) \cite{bender3}. Another example is that they were used in the construction of an angle operator for time dependent oscillators \cite{angle} and of a set of eigenfunctions and eigenvalues for a particular system of coupled nonlinear integro-differential equations describing the ground state of a one-dimensional Fermi gas \cite{eig}. In the study of non-Hermitian Hamiltonians, the basis operators played a major role in obtaining the Hermitian counterpart of a particular Hamiltonian \cite{nonherm} and the operator which is needed to define a positive-definite inner product of the Hilbert space where a $\mathcal{PT}$ symmetric Hamiltonian acts on \cite{ptsym1, ptsym2}. These operators were also applied in the integration of the Heisenberg operator equations for various potentials i.e. anharmonic potentials up to order four and inverse-power law potentials \cite{heis1, heis2, heis3}. Moreover, the properties of the $\opr{T}_{m,n}$'s were shown in a different light by considering them to be components of an SU(2) tensor operator \cite{tensorop}, while the spectral properties of a particular set of these operators for spatially confined particle has also been obtained in \cite{spectr}.

The paper is organized as follows. In Section-\ref{proof1} we work out the proof for the operator $\opr{T}_{-1,1}$ to concretely demonstrate the method used in proving the general result. In Section-\ref{proof2} we give the proof for the general case  $\opr{T}_{-m,n}$. In Section-\ref{conc} we conclude. Appendices are provided to detail some aspects of the proof in the general case.

\section{The proof for $\opr{T}_{-1,1}$}\label{proof1}
For concreteness, we first show that the formal operator $\opr{T}_{-1,1}$, in particular the integral operator
\begin{equation}\label{toa}
\left(\opr{T}_{-1,1} \varphi \right)\!(q) =  \int_{-\infty}^{\infty}(q+q')\mbox{sgn}(q-q')\varphi (q') dq',
\end{equation}
has a non-trivial dense domain in the Hilbert space $L^2(\mathbb{R})$, thereby proving that $\opr{T}_{-1,1}$ is a densely defined Hilbert space operator. The operator $\opr{T}_{-1,1}$ is directly proportional to the free time of arrival operator \cite{freetoa,galapon3,galapon8}.
Let us assume that $\opr{T}_{-1,1}$ has a non-trivial domain, $D(\opr{T}_{-1,1})\subseteq L^2(\mathbb{R})$. It is now our objective to show that $D(\opr{T}_{-1,1})$ exists and is dense. If  $\varphi(q)$ belongs to $D(\opr{T}_{-1,1})$, then $(\opr{T}_{-1,1}\varphi)(q)$ is necesarily square integrable, 
\begin{equation}\label{sqr}
\int_{-\infty}^{\infty}|(\opr{T}_{-1,1} \varphi )(q)|^{2}dq<\infty.
\end{equation} 
Certainly $D(\opr{T}_{-1,1})$ is not the entire Hilbert space because $\opr{T}_{-1,1}$ acting on $e^{-q^2}\in L^2({\mathbb{R}})$ yields a function whose behavior at infinity is $\int_{-\infty}^{\infty}(q+q')\mbox{sgn}(q-q') e^{-q'^2} \mbox{d}q' \sim \sqrt{\pi} \, q$ as $\abs{q}\rightarrow\infty$, which is not square integrable in the real line; but it is neither empty because the vector $(1-2q)e^{-q^2}\in L^2(\mathbb{R})$ belongs to the domain because $\int_{-\infty}^{\infty}(1-2q'^2)e^{-q'^2}(q+q')\mbox{sgn}(q-q')\mbox{d}q'=(1+4q^2)e^{-q^2}$, which is square integrable.

Let us now identify the necessary and sufficient conditions in order for the square integrability condition \ref{sqr} to be satisfied. To proceed let us expand equation \ref{toa} using the identity $\mbox{sgn}(x)=H(x)-H(-x)$, where $H(x)$  is the Heaviside step function. The expansion yields
\begin{eqnarray}\label{expandtoa}
 \left(\opr{T}_{-1,1} \varphi \right)\!\!(q) =  q\left(\int_{-\infty}^{q}\!\!\!\!\varphi (q') dq'\!\! -\!\! \int_{q}^{\infty}\!\!\!\!\!\!\varphi (q') dq'\right) \!\!+\! \! \left(\int_{-\infty}^{q}\!\!\!\!\!q' \varphi (q') dq' \!\!-\!\! \int_{q}^{\infty}\!\!\!\!\! q' \varphi (q') dq'\right) .
\end{eqnarray}
A necessary but not sufficient condition for $(\opr{T}_{-1,1}\varphi)\!(q)$ to be square integrable is that the right hand side of equation \ref{expandtoa} vanishes  as $|q| \rightarrow \infty$. This implies the following necessary integral conditions 
\begin{equation}\label{zerocond}
\int_{-\infty}^{\infty}\varphi (q) dq = 0, \;\;\;\;\; \int_{-\infty}^{\infty}q\varphi (q) dq = 0 .
\end{equation}
A sufficient condition to ensure that $\varphi(q)$ belongs to the domain is that  each term in the right hand side of \ref{expandtoa} is square integrable at infinity. This is guaranteed provided the following integrals have the indicated minimum rate of decrease for arbitrarily large $|q|$,
\begin{equation}\label{asymcond}
\left|\int_{|q|}^{\infty}\varphi (\pm q') dq'\right| = O(|q|^{-2}), \;\;\;\;\; \left|\int_{|q|}^{\infty}q'\varphi (\pm q') dq'\right| = O(|q|^{-1}), |q|\rightarrow\infty .
\end{equation}
The integrals may decrease faster, but not slower. Under these conditions, the two terms are square integrable in the real line. Once the conditions Eqs.~(\ref{asymcond}) are satisfied, then so are Eqs.~(\ref{zerocond}). 

Let us now construct a complete set of vectors, $\mathcal{S}(\opr{T}_{-1,1})$, of $L^2(\mathbb{R})$ that is mapped by $\opr{T}_{-1,1}$ into $L^2(\mathbb{R})$.   Elements of $S(\opr{T}_{-1,1})$ must necessarily satisfy the integral conditions \ref{zerocond}.  Since the conditions are a pair, we need at least a pair of linearly independent vectors and their linear sum to ensure \ref{zerocond}. Let $\varphi(q) = \alpha \varphi_{1}(q) + \beta \varphi_{2}(q)$ belong to $S(\opr{T}_{-1,1})$, where $\varphi_{1}(q)$ and $\varphi_{2}(q)$ are linearly independent functions, and $\alpha$ and $\beta$ are constants such that conditions \ref{zerocond} are satisfied. Substituting $\varphi(x)$ to Eqs.~(\ref{zerocond}), we obtain the following matrix expression for $\alpha$ and $\beta$,
\[ \left[ \begin{array}{cc}
\int_{-\infty}^{\infty}\varphi_{1}(q)dq & \int_{-\infty}^{\infty}\varphi_{2}(q)dq\\
\int_{-\infty}^{\infty}y\varphi_{1}(q)dq & \int_{-\infty}^{\infty}y\varphi_{2}(q)dq
\end{array} \right]
\left[ \begin{array}{c}
\alpha \\
\beta
\end{array} \right] = 0
\] 
A solution for $\alpha$ and $\beta$ exist if the determinant of the matrix of the coefficients vanishes. This gives the following condition for the integrals of $\varphi_{1}(q)$ and $\varphi_{2}(q)$,
\begin{equation}\label{constcond}
\left[\int_{-\infty}^{\infty}\varphi_{1}(q)dq\right]\left[\int_{-\infty}^{\infty}q\varphi_{2}(q)dq\right] = \left[\int_{-\infty}^{\infty}\varphi_{2}(q)dq\right]\left[\int_{-\infty}^{\infty}q\varphi_{1}(q)dq\right]
\end{equation}
Eq.~(\ref{constcond}) can be satisfied by letting $\varphi_{1}(x)$ and $\varphi_{2}(x)$ have definite parities. That is, either both of them are even or both are odd. With these conditions, we can write $\varphi^{(1)}(q) = \left(\int_{-\infty}^{\infty}\varphi_{1}(y)dy\right)\varphi_{2}(q) - \left(\int_{-\infty}^{\infty}\varphi_{2}(y)dy\right)\varphi_{1}(q)$ for even $\varphi_{1}(x)$ and $\varphi_{2}(x)$; and $\varphi^{(2)}(x) = \left(\int_{-\infty}^{\infty}y\varphi_{1}(y)dy\right)\varphi_{2}(x) - \left(\int_{-\infty}^{\infty}y\varphi_{2}(y)dy\right)\varphi_{1}(x)$ for odd $\varphi_{1}(x)$ and $\varphi_{2}(x)$. By direct substitution, one can see that they satisfy Eqs.~(\ref{zerocond}) provided that the integrals exist. The vectors $\varphi^{(1)}(q)$ and $\varphi^{(2)}(q)$ must satisfy the minimum asymptotic conditions \ref{asymcond}.

 We now choose a complete set of vectors in the Hilbert space $L^2(\mathbb{R})$ with definite parities. The set of vectors $\{\psi_{k}(q) = (2^{k/2}\sqrt{k!}\sqrt[4]{\pi})^{-1}H_{k}(q)\mbox{e}^{-q^{2}/2}, k=0,1,\dots\}$, where the $H_{k}(q)$'s are the Hermite polynomials, is complete in  $L^2(\mathbb{R})$. Each $\psi_k(q)$ has a definite parity, even (odd) for even (odd) $k$.  We can then define the basic vectors satisfying Eqs.~(\ref{zerocond}) as
\begin{eqnarray}\label{basicvec}
\psi_{m,n}^{(1)}(q) &=& A_{n}^{(1)}\psi_{2m}(q)-A_{m}^{(1)}\psi_{2n}(q), \\
\psi_{m,n}^{(2)}(q) &=& A_{n}^{(2)}\psi_{2m+1}(q)-A_{m}^{(2)}\psi_{2n+1}(q),
\end{eqnarray}
for all natural numbers $m$ and $n$ with $m \neq n$. The coefficients are given by 
\begin{eqnarray}
A_{n}^{(1)} &=& \int_{-\infty}^{\infty}\psi_{2n}(y)dy = \frac{\sqrt{2}\sqrt[4]{\pi}\sqrt{(2n)!}}{2^{n}n!},\\
A_{n}^{(2)} &=& \int_{-\infty}^{\infty}y\psi_{2n+1}(y)dy = \frac{2\sqrt[4]{\pi}\sqrt{(2n+1)!}}{2^{n}n!}.
\end{eqnarray}
 These coefficients have been obtained using tabulated integrals \cite{tabl}.
It can be shown that the vectors $(\opr{T}_{-1,1}\psi_{m,n}^{(1)})(q)$ and $(\opr{T}_{-1,1}\psi_{m,n}^{(2)})(q)$ exponentially decay as $|q|\rightarrow \infty$ so that Eqs.~(\ref{asymcond}) are satisfied. We have constructed the subset $S(\opr{T}_{-1,1})=\{\psi^{(1)}_{m,n}(q),\psi^{(2)}_{m',n'}(q); m,m', n, n' = 0, 1,2,\dots\}$

We now show that $S(\opr{T}_{-1,1})$ is a complete set (in fact over complete). A sequence of vectors is complete if the only vector orthogonal to every element of the sequence is the zero vector.  Let $n$ be fixed and consider the sequence of vectors $\{\psi_{m,n}^{(1)},\psi_{m,n}^{(2)}, m=0,1,\dots\}$. This sequence is complete if $<\psi_{m,n}^{(k)}|\phi>=0$  for all $m=0,1,2,\dots$ and $k=1,2$ implies that $\phi=0$. 
Since the vectors $\psi_{k}(x)$ form a complete set of orthonormal vectors, we can write $\phi(x) = \sum_{k=0}^{\infty}\phi_{k}\psi_{k}(x)$. We must now show that $\phi_{k}=0$ for all $k$ so that $\phi(x)=0$. 
The orthonormality of the $\psi_k(q)$'s and the conditions $<\psi_{m,n}^{(1)}|\phi> = 0$ and $<\psi_{m,n}^{(2)}|\phi> = 0$ for all $m$ imply
\begin{eqnarray}\label{coeffbehaviour}
\phi_{2m} &=& \left(\frac{\sqrt{(2n)!}}{2^{n}n!}\right)^{-1}\left(\frac{\sqrt{(2m)!}}{2^{m}m!}\right)\phi_{2n}, \\
 \phi_{2m+1} &=& \left(\frac{\sqrt{(2n+1)!}}{2^{n}n!}\right)^{-1}\left(\frac{\sqrt{(2m+1)!}}{2^{m}m!}\right)\phi_{2n+1} .
\end{eqnarray}
Let us consider the behavior of the coefficients for $m>n$ as $m\rightarrow\infty$. Using Stirling's formula: $m! \sim \sqrt{2\pi}m^{m+1/2}\mbox{e}^{-m}$ as $m\rightarrow\infty$, we obtain the following asymptotic behaviors for the coefficients:
\begin{eqnarray}
\phi_{2m} &\sim& \left(\frac{\sqrt{(2n)!}}{2^{n}n!}\right)^{-1}\frac{1}{m^{1/4}}\phi_{2n}, \label{asym1}\\
 \phi_{2m+1} &\sim& \left(\frac{\sqrt{(2n+1)!}}{2^{n}n!}\right)^{-1}m^{1/4}\phi_{2n+1}\label{asym2}
\end{eqnarray}

Since $\phi(q)$ should belong to $L^2(\mathbb{R})$, it is necessary that $\sum_{k=0}^{\infty}|\phi_{k}|^{2}<\infty$. We see immediately that from Eqs.~(\ref{asym2}), $\phi_{2m+1}$ cannot satisfy this for non-zero $\phi_{2n+1}$ since $\phi_{2m+1}$ algebraically diverges as $m\rightarrow\infty$. It is also not possible for the coefficients $\phi_{2m}$ in \ref{asym1}, since they behave as $\phi_{2m} \sim \mathcal{O}(m^{-1/4})$ as $m\rightarrow\infty$ for non-zero $\phi_{2n}$. It is then necessary to have $\phi_{2n+1}=0$ and $\phi_{2n}=0$. From Eqs.~(\ref{coeffbehaviour}), we see that $\phi_{2m+1}=0$ and $\phi_{2m}=0$ for any $m \neq n$.  We then have $\phi_{k}=0$ for any $k$. That is, the only vector orthogonal to the sequence for a given $n$ is the zero vector. Hence the sequence  $\{\psi_{m,n}^{(1)},\psi_{m,n}^{(2)}, m=0,1,\dots\}$ is complete in $L^2(\mathbb{R})$. Also since $n$ is arbitrarily chosen, our conclusion should hold for any $n$, so that for each natural number $n$ we have a complete set. Then the set $S(\opr{T}_{-1,1})$ is (over) complete. 

We can now assign the linear span of $S(\opr{T}_{-1,1})$ to be the domain $D(\opr{T}_{-1,1})$ of $\opr{T}_{-1,1}$. Since $S(\opr{T}_{-1,1})$ is a complete set, its linear span $D(\opr{T}_{-1,1})$  is necessarily dense in $L^2(\mathbb{R})$. Hence the operator $\opr{T}_{-1,1}:D(\opr{T}_{-1,1})\subset L^2(\mathbb{R})\mapsto L^2(\mathbb{R})$ is a densely defined operator.

\section{The proof for $\opr{T}_{-m,n}$}\label{proof2}
\subsection{The necessary and sufficient conditions}
Let us first identify the necessary and sufficient conditions for a vector in $L^2(\mathbb{R})$  to belong in the domain of $\opr{T}_{-m,n}$. For convenience, we shift $m$ to $m+1$. We first expand the kernel to assume the form
\begin{eqnarray}
\left<q|\opr{T}_{-(m+1),n}|q'\right> &=&  C_{m,n}(q+q')^{n}(q-q')^{m}\mbox{sgn}(q-q')  \nonumber\\
&=& \sum_{\ell=0}^{N}\sum_{k=0}^{\ell}D_{m,n,\ell,k}\;\;q^{N-\ell}q'^{\ell} \mbox{sgn}(q-q')  \nonumber
\end{eqnarray}
where, $N=n+m=1,2,3,\dots$, $D_{m,n,\ell,k} = (-1)^{k}\left(^{\;\;n}_{\ell-k}\right)\left(^{m}_{k}\right)C_{m,n}$, and the $C_{m,n} = \frac{i(-1)^{m/2}}{2^{n+1}\hbar^{m+1}m!}$ are constants that are not relevant to the proof. 
The action of $\opr{T}_{-(m+1),n}$ in position representation is then given by
\begin{eqnarray}
&& \left(\opr{T}_{-(m+1),n}\varphi\right)(q) = \int_{\infty}^{\infty} \left<q|\opr{T}_{-(m+1),n}|q'\right>\varphi(q')dq'  \nonumber\\
&& \;\;\;\;\;\;\;\;\; = \sum_{\ell=0}^{N}\sum_{k=0}^{\ell}D_{m,n,\ell,k}\left(q^{N-\ell}\int_{\infty}^{\infty}q'^{\ell} \mbox{sgn}(q-q')\varphi(q')dq'\right)
\nonumber
\end{eqnarray}
Again we must have the square integrability condition $\int_{-\infty}^{\infty}|(\opr{T}_{-(m+1),n}\varphi)(q)|^{2}dq<\infty$, which implies that $(\opr{T}_{-(m+1),n}\varphi)\!(q) \rightarrow 0$ as $|q|\rightarrow \infty$. Consider expanding the integral in the parenthesis to get $q^{N-\ell}\left(\int_{-\infty}^{q}q'^{\ell} \varphi(q') dq' - \int_{q}^{\infty}q'^{\ell} \varphi(q') dq'\right)$. The vanishing of $(\opr{T}_{-(m+1),n}\varphi)\!(q)$ at infinity requires the vanishing of the following integrals,
\begin{equation}\label{cond}
\int_{-\infty}^{\infty}q'^{\ell}\varphi(q')dq' = 0, \;\;\; \ell=0,1,\dots, N .
\end{equation}
Strictly speaking, the integrals $\int_{-\infty}^{q}q'^{\ell} \varphi(q') dq' - \int_{q}^{\infty}q'^{\ell} \varphi(q') dq'$ must vanish faster than any integer power of $|q^{-1}|$. That is, we should have
\begin{equation}\label{asymptoticcond}
\left| \int_{|q|}^{\infty}q'^{\ell}\varphi(\pm q')dq' \right| \sim O\left(\frac{1}{|q|^{N-\ell +1}}\right), \;\;\; |q|\rightarrow \infty .
\end{equation}

\subsection{The basic vectors} 
We now write the complete set of basic vectors, $S(\opr{T}_{-(m+1),n})$, so that the stated conditions are satisfied. Eqs (\ref{cond}) are $N+1$ conditions for the function $\varphi(q) \in S(\opr{T}_{-(m+1),n})$ and much like in the previous section, we need $N+1$ linearly independent vectors and their linear sum to ensure the conditions Eqs (\ref{cond}) are satisfied. We choose the $\psi_{k}$'s to be the linearly independent vectors and linear sums (the basic vectors) are as follows:
\begin{eqnarray}\label{basicv}
&& \varphi^{(e)}(q) = \sum_{i=0}^{N}\alpha_{i}^{(e)}\psi_{2k_{i}^{(e)}}(q)  \nonumber\\
&& \varphi^{(o)}(q) = \sum_{i=0}^{N}\alpha_{i}^{(o)}\psi_{2k_{i}^{(o)}+1}(q)  \nonumber\\
\end{eqnarray}
where,

\begin{displaymath}
\alpha_{i}^{(e\backslash o)} = (-1)^{i}
\left|
\begin{array}[pos]{ccccccc}
  A_{0,0}^{(e \backslash o)} & A_{0,1}^{(e\backslash o)} & \cdots & A_{0,i-1}^{(e\backslash o)} & A_{0,i+1}^{(e\backslash o)}  & \cdots & A_{0,N}^{(e\backslash o)} \\
  A_{1,0}^{(e \backslash o)} & A_{1,1}^{(e\backslash o)} &  \cdots & A_{1,i-1}^{(e\backslash o)} & A_{1,i+1}^{(e\backslash o)} & \cdots & A_{1,N}^{(e\backslash o)} \\
  A_{2,0}^{(e \backslash o)} & A_{2,1}^{(e\backslash o)} &  \cdots & A_{2,i-1}^{(e\backslash o)} & A_{2,i+1}^{(e\backslash o)} & \cdots & A_{2,N}^{(e\backslash o)} \\
	\vdots & \vdots & \ddots & \vdots & \vdots & \ddots & \vdots \\
  A_{N-1,0}^{(e \backslash o)} & A_{N-1,1}^{(e\backslash o)} & \cdots & A_{N-1,i-1}^{(e\backslash o)} & A_{N-1,i+1}^{(e\backslash o)} & \cdots & A_{N-1,N}^{(e\backslash o)} \\
\end{array}
\right|
\end{displaymath}
\begin{eqnarray}
&& A_{j,i}^{(e)} = \int_{-\infty}^{\infty} q^{2j}\psi_{2k_{i}^{(e)}}(q)dq \nonumber\\
&& A_{j,i}^{(o)} = \int_{-\infty}^{\infty} q^{2j+1}\psi_{2k_{i}^{(o)}+1}(q)dq \nonumber\\
&& \psi_{n}(q) = \frac{1}{\sqrt{2^{n}n!\sqrt{\pi}}}H_{n}(q)\exp\left(-q^{2}/2\right) \nonumber\\
\nonumber
\end{eqnarray}
and $k_{i}^{(e \backslash o)} \neq k_{j \neq i}^{(e \backslash o)}$ are integers. This is to ensure the linear independence of the terms in equations (\ref{basicv}). Since the $\alpha_{i}^{(e \backslash o)}$'s can be interpreted as cofactors of a matrix, we can rewrite the basic vectors equations (\ref{basicv}) as determinants of the aforementioned matrix.
\begin{eqnarray}\label{evendet}
&& \varphi^{(e)}(q) = 
\left|
\begin{array}[pos]{ccccc}
  \psi_{2k_{0}^{(e)}}(q) & \psi_{2k_{1}^{(e)}}(q) & \psi_{2k_{2}^{(e)}}(q) & \cdots & \psi_{2k_{N}^{(e)}}(q) \\
  A_{0,0}^{(e)} & A_{0,1}^{(e)} & A_{0,2}^{(e)} & \cdots & A_{0,N}^{(e)}  \\
  A_{1,0}^{(e)} & A_{1,1}^{(e)} & A_{1,2}^{(e)} & \cdots & A_{1,N}^{(e)}  \\
  A_{2,0}^{(e)} & A_{2,1}^{(e)} & A_{2,2}^{(e)} & \cdots & A_{2,N}^{(e)}  \\
	\vdots & \vdots & \vdots & \ddots & \vdots \\
  A_{N-1,0}^{(e)} & A_{N-1,1}^{(e)} & A_{N-1,2}^{(e)} & \cdots & A_{N-1,N}^{(e)}  \\
\end{array}
\right| \nonumber \\
&& = 
\left|
\begin{array}[pos]{ccc}
  \psi_{2k_{0}^{(e)}}(q)  & \cdots & \psi_{2k_{N}^{(e)}}(q) \\
  \int_{-\infty}^{\infty}\psi_{2k_{0}^{(e)}}(q)dq & \cdots & \int_{-\infty}^{\infty}\psi_{2k_{N}^{(e)}}(q)dq  \\
  \int_{-\infty}^{\infty}q^{2}\psi_{2k_{0}^{(e)}}(q)dq & \cdots & \int_{-\infty}^{\infty}q^{2}\psi_{2k_{N}^{(e)}}(q)dq  \\
  \int_{-\infty}^{\infty}q^{4}\psi_{2k_{0}^{(e)}}(q)dq & \cdots & \int_{-\infty}^{\infty}q^{4}\psi_{2k_{N}^{(e)}}(q)dq  \\
	\vdots & \ddots & \vdots \\
  \int_{-\infty}^{\infty}q^{2(N-1)}\psi_{2k_{0}^{(e)}}(q)dq & \cdots & \int_{-\infty}^{\infty}q^{2(N-1)}\psi_{2k_{N}^{(e)}}(q)dq  \\  
\end{array}
\right| \nonumber \\
&& = |\mathsf{\Psi}^{(e)}| \nonumber\\
\end{eqnarray}
\begin{eqnarray}\label{odddet}
&& \varphi^{(o)}(q) = 
\left|
\begin{array}[pos]{ccccc}
  \psi_{2k_{0}^{(o)}+1}(q) & \psi_{2k_{1}^{(o)}+1}(q) & \psi_{2k_{2}^{(o)}+1}(q) & \cdots & \psi_{2k_{N}^{(o)}+1}(q) \\
  A_{0,0}^{(o)} & A_{0,1}^{(o)} & A_{0,2}^{(o)} & \cdots & A_{0,N}^{(o)}  \\
  A_{1,0}^{(o)} & A_{1,1}^{(o)} & A_{1,2}^{(o)} & \cdots & A_{1,N}^{(o)}  \\
  A_{2,0}^{(o)} & A_{2,1}^{(o)} & A_{2,2}^{(o)} & \cdots & A_{2,N}^{(o)}  \\
	\vdots & \vdots & \vdots & \ddots & \vdots \\
  A_{N-1,0}^{(o)} & A_{N-1,1}^{(o)} & A_{N-1,2}^{(o)} & \cdots & A_{N-1,N}^{(o)}  \\
\end{array}
\right| \nonumber \\
&& = 
\left|
\begin{array}[pos]{ccc}
  \psi_{2k_{0}^{(o)}+1}(q) &  \cdots & \psi_{2k_{N}^{(o)}+1}(q) \\
  \int_{-\infty}^{\infty}q\psi_{2k_{0}^{(o)}+1}(q)dq & \cdots & \int_{-\infty}^{\infty}q\psi_{2k_{N}^{(o)}+1}(q)dq  \\
  \int_{-\infty}^{\infty}q^{3}\psi_{2k_{0}^{(o)}+1}(q)dq & \cdots & \int_{-\infty}^{\infty}q^{3}\psi_{2k_{N}^{(o)}+1}(q)dq  \\
  \int_{-\infty}^{\infty}q^{5}\psi_{2k_{0}^{(o)}+1}(q)dq & \cdots & \int_{-\infty}^{\infty}q^{5}\psi_{2k_{N}^{(o)}+1}(q)dq  \\
	\vdots & \ddots & \vdots \\
  \int_{-\infty}^{\infty}q^{2(N-1)+1}\psi_{2k_{0}^{(o)}+1}(q)dq & \cdots & \int_{-\infty}^{\infty}q^{2(N-1)+1}\psi_{2k_{N}^{(o)}+1}(q)dq  \\
\end{array}
\right| \nonumber \\
&& = |\mathsf{\Psi}^{(o)}| \nonumber\\
\end{eqnarray}
where $\mathsf{\Psi}^{(e \backslash o)}$ are $(N+1) \times (N+1)$ matrices. Using the cofactor expansion along the first row, it is easy to show that Eqs (\ref{evendet}-\ref{odddet}) are equivalent to equations (\ref{basicv}) with the $\alpha_{i}$'s as the cofactors $\alpha_{i}^{(e\backslash o)} = \mathsf{C}_{0,i}[\mathsf{\Psi}^{(e \backslash o)}]$. 

We now show explicitly that Eqs (\ref{evendet}-\ref{odddet}), or equivalently, Eqs (\ref{basicv}) indeed satisfy the conditions equations (\ref{cond}). Using the cofactor expansion along the first row of Eq (\ref{evendet}), we test if $\int_{-\infty}^{\infty}q^{2j'+1}\varphi^{(e)}(q)dq = 0$ where, $\ell = 2j'+1 = 1,3,5,... ,N_{odd}$ and $N_{odd}$ is the greatest odd integer satisfying $N_{odd}\leq N$.
\begin{displaymath}
\int_{-\infty}^{\infty}q^{2j'+1}\varphi^{(e)}(q)dq = \sum_{i=0}^{N}\mathsf{C}_{0,i}[\mathsf{\Psi}^{(e)}]\int_{-\infty}^{\infty}q^{2j'+1}\psi_{2k_{i}^{(e)}}(q)dq = 0
\end{displaymath}
since $\int_{-\infty}^{\infty}q^{2j'+1}\psi_{2k_{i}^{(e)}}(q)dq = 0$. That is, since $q^{2j'+1}\psi_{2k_{i}^{(e)}}(q)$ is an odd function. Next, we test
\begin{eqnarray}
&& \int_{-\infty}^{\infty}q^{2j'}\varphi^{(e)}(q)dq = \sum_{i=0}^{N}\mathsf{C}_{0,i}[\mathsf{\Psi}^{(e)}]\int_{-\infty}^{\infty}q^{2j'}\psi_{2k_{i}^{(e)}}(q)dq  \nonumber\\
&& = 
\left|
\begin{array}[pos]{ccc}
  \int_{-\infty}^{\infty}q^{2j'}\psi_{2k_{0}^{(e)}}(q)dq & \cdots & \int_{-\infty}^{\infty}q^{2j'}\psi_{2k_{N}^{(e)}}(q)dq \\
  \int_{-\infty}^{\infty}\psi_{2k_{0}^{(e)}}(q)dq & \cdots & \int_{-\infty}^{\infty}\psi_{2k_{N}^{(e)}}(q)dq  \\
  \int_{-\infty}^{\infty}q^{2}\psi_{2k_{0}^{(e)}}(q)dq & \cdots & \int_{-\infty}^{\infty}q^{2}\psi_{2k_{N}^{(e)}}(q)dq  \\
  \int_{-\infty}^{\infty}q^{4}\psi_{2k_{0}^{(e)}}(q)dq &  \cdots & \int_{-\infty}^{\infty}q^{4}\psi_{2k_{N}^{(e)}}(q)dq  \\
	\vdots & \ddots & \vdots \\
  \int_{-\infty}^{\infty}q^{2(N-1)}\psi_{2k_{0}^{(e)}}(q)dq & \cdots & \int_{-\infty}^{\infty}q^{2(N-1)}\psi_{2k_{N}^{(e)}}(q)dq  \\  
\end{array}
\right| = 0 \nonumber \\
\nonumber
\end{eqnarray}
where, $\ell = 2j' = 0,2,4,... ,N_{even}$ and $N_{even}$ is the greatest even integer satisfying $N_{even}\leq N$. The determinant vanishes since the first row is always a copy of another row in the determinant. We then have $\int_{-\infty}^{\infty}q^{\ell}\varphi^{(e)}(q)dq = 0$, $\ell=0,1,2,...,N$. That is, the $\varphi^{(e)}$'s satisfy the conditions equations (\ref{cond}).

Using a similar procedure, we test if $\int_{-\infty}^{\infty}q^{2j'}\varphi^{(o)}(q)dq = 0$ where, $\ell = 2j' = 0,2,4,... ,N_{even}$ and $N_{even}$ is the greatest even integer satisfying $N_{even}\leq N$.
\begin{displaymath}
\int_{-\infty}^{\infty}q^{2j'}\varphi^{(o)}(q)dq = \sum_{i=0}^{N}\mathsf{C}_{0,i}[\mathsf{\Psi}^{(o)}]\int_{-\infty}^{\infty}q^{2j'}\psi_{2k_{i}^{(o)}+1}(q)dq = 0
\end{displaymath}
since $\int_{-\infty}^{\infty}q^{2j'}\psi_{2k_{i}^{(o)}+1}(q)dq = 0$. That is, since $q^{2j'}\psi_{2k_{i}^{(o)}+1}(q)$ is an odd function. Next, we test
\begin{eqnarray}
&& \int_{-\infty}^{\infty}q^{2j'+1}\varphi^{(o)}(q)dq = \sum_{i=0}^{N}\mathsf{C}_{0,i}[\mathsf{\Psi}^{(o)}]\int_{-\infty}^{\infty}q^{2j'+1}\psi_{2k_{i}^{(o)}+1}(q)dq  \nonumber\\
&& = 
\left|
\begin{array}[pos]{ccc}
  \int_{-\infty}^{\infty}q^{2j'+1}\psi_{2k_{0}^{(o)}+1}(q)dq & \cdots & \int_{-\infty}^{\infty}q^{2j'+1}\psi_{2k_{N}^{(o)}+1}(q)dq \\
  \int_{-\infty}^{\infty}q\psi_{2k_{0}^{(o)}+1}(q)dq & \cdots & \int_{-\infty}^{\infty}q\psi_{2k_{N}^{(o)}+1}(q)dq  \\
  \int_{-\infty}^{\infty}q^{3}\psi_{2k_{0}^{(o)}+1}(q)dq & \cdots & \int_{-\infty}^{\infty}q^{3}\psi_{2k_{N}^{(o)}+1}(q)dq  \\
  \int_{-\infty}^{\infty}q^{5}\psi_{2k_{0}^{(o)}+1}(q)dq & \cdots & \int_{-\infty}^{\infty}q^{5}\psi_{2k_{N}^{(o)}+1}(q)dq  \\
	\vdots & \ddots & \vdots \\
  \int_{-\infty}^{\infty}q^{2(N-1)+1}\psi_{2k_{0}^{(o)}+1}(q)dq  & \cdots & \int_{-\infty}^{\infty}q^{2(N-1)+1}\psi_{2k_{N}^{(o)}+1}(q)dq  \\
\end{array}
\right| = 0  \nonumber \\ 
\nonumber
\end{eqnarray}
where, $\ell = 2j'+1 = 1,3,5,... ,N_{odd}$ and $N_{odd}$ is the greatest odd integer satisfying $N_{odd}\leq N$. The determinant vanishes since the first row is always a copy of another row in the determinant. We then have $\int_{-\infty}^{\infty}q^{\ell}\varphi^{(o)}(q)dq = 0$, $\ell=0,1,2,...,N$. That is, the $\varphi^{(o)}$'s also satisfy the conditions Eqs (\ref{cond}).
It can also be seen that since the basic vectors are just linear combinations of $\psi_{n}(q) \propto H_{n}(q)\exp\left(-q^{2}/2\right)$, then Eqs (\ref{basicv}) indeed satisfy Eqs (\ref{asymptoticcond}) using the behavior $\int_{|x|}^{\infty}\exp\left(-y^{2}/2\right)y^{k}dy \sim |x|^{k-1}\exp\left(-x^{2}/2\right)$ as $|x|\rightarrow \infty$. Then, equations (\ref{basicv}) indeed represent the basic vectors $\varphi^{(e \backslash o)}(q) \in S(\opr{T}_{-(m+1),n})$. 

\subsection{Completeness of the basic vectors}
We now show that $S(\opr{T}_{-(m+1),n})$ is indeed an (over) complete set. If the basic vectors $\varphi^{(e \backslash o)}(q)$ form a complete set, then the only vector orthogonal to every element of the set is the zero vector. In symbols, if the equalities $<\phi |\varphi^{(e \backslash o)}> = 0$ hold for all $\varphi^{(e \backslash o)} \in S(\opr{T}_{-(m+1),n})$ and for $\phi \in L^{2}(\mathbb{R})$, then this implies that $\phi=0$. Since the $\psi_{n}$'s form a basis in $L^{2}(\mathbb{R})$, we have $\phi(q) = \sum_{n=0}^{\infty}\phi_{n}^{(e)}\psi_{2n}(q) + \sum_{n=0}^{\infty}\phi_{n}^{(o)}\psi_{2n+1}(q)$ where, $\phi_{n}^{(e)} = <\psi_{2n}|\phi>$ and $\phi_{n}^{(o)} = <\psi_{2n+1}|\phi>$. Equivalently, if $S(\opr{T}_{-(m+1),n})$ is indeed a complete set, and if $\sum_{n=0}^{\infty}|\phi_{n}^{(e)}|^{2} + \sum_{n=0}^{\infty}|\phi_{n}^{(o)}|^{2} < \infty$, then we should have $\phi_{n}^{(e)} = 0 = \phi_{n}^{(o)}$ for all $n$. Using equations (\ref{basicv}), we can write
\begin{eqnarray}
&& <\phi|\varphi^{(e)}> = 0 = \sum_{i=0}^{N}\mathsf{C}_{0,i}[\mathsf{\Psi}^{(e)}] \int_{-\infty}^{\infty}\phi^{*}(q)\psi_{2k_{i}^{(e)}}(q)dq \nonumber\\
&& <\phi|\varphi^{(o)}> = 0 = \sum_{i=0}^{N}\mathsf{C}_{0,i}[\mathsf{\Psi}^{(o)}] \int_{-\infty}^{\infty}\phi^{*}(q)\psi_{2k_{i}^{(o)}+1}(q)dq \nonumber\\
\nonumber\\
\end{eqnarray}
or, more compactly by using equations (\ref{evendet}) and (\ref{odddet}),
\begin{eqnarray}\label{detvanish}
&& <\phi|\varphi^{(e \backslash o)}> = 0 = \left|
\begin{array}[pos]{ccccc}
  \phi_{k_{0}}^{(e \backslash o)*} & \phi_{k_{1}}^{(e \backslash o)*} & \phi_{k_{2}}^{(e \backslash o)*} & \cdots & \phi_{k_{N}}^{(e \backslash o)*} \\
  A_{0,0}^{(e \backslash o)} & A_{0,1}^{(e \backslash o)} & A_{0,2}^{(e \backslash o)} & \cdots & A_{0,N}^{(e \backslash o)}  \\
  A_{1,0}^{(e \backslash o)} & A_{1,1}^{(e \backslash o)} & A_{1,2}^{(e \backslash o)} & \cdots & A_{1,N}^{(e \backslash o)}  \\
  A_{2,0}^{(e \backslash o)} & A_{2,1}^{(e \backslash o)} & A_{2,2}^{(e \backslash o)} & \cdots & A_{2,N}^{(e \backslash o)}  \\
	\vdots & \vdots & \vdots & \ddots & \vdots \\
  A_{N-1,0}^{(e \backslash o)} & A_{N-1,1}^{(e \backslash o)} & A_{N-1,2}^{(e \backslash o)} & \cdots & A_{N-1,N}^{(e \backslash o)}  \\
\end{array}
\right|\nonumber \\
\nonumber\\
\end{eqnarray}
where the $k_{i}$'s are understood to be the $k_{i}^{(e)}$'s for $\phi^{(e)}$ and $A_{j,i}^{(e)}$, and the $k_{i}^{(o)}$'s for $\phi^{(o)}$ and $A_{j,i}^{(o)}$. The vanishing of the determinant in Eq (\ref{detvanish}) suggests that we can relate the $\phi_{n}^{(e \backslash o)*}$'s. We let the set of $N$ indices $\{k_{1}, k_{2}, ..., k_{N}\}$ ($k_{i} \neq k_{j \neq i}$), and in turn, the set of $N$ coefficients $\{\phi_{k_{1}}^{(e \backslash o)*}, \phi_{k_{2}}^{(e \backslash o)*}, ..., \phi_{k_{N}}^{(e \backslash o)*} \}$ to be fixed but arbitrary so that $\phi_{k_{0}}^{(e \backslash o)*}$ depends on the others by
\begin{equation}\label{phirel}
-\phi_{k_{0}}^{(e \backslash o)*}\mathsf{C}_{0,0}[\mathsf{\Psi}^{(e \backslash o)}] = \left|
\begin{array}[pos]{ccccc}
  0 & \phi_{k_{1}}^{(e \backslash o)*} & \phi_{k_{2}}^{(e \backslash o)*} & \cdots & \phi_{k_{N}}^{(e \backslash o)*} \\
  A_{0,0}^{(e \backslash o)} & A_{0,1}^{(e \backslash o)} & A_{0,2}^{(e \backslash o)} & \cdots & A_{0,N}^{(e \backslash o)}  \\
  A_{1,0}^{(e \backslash o)} & A_{1,1}^{(e \backslash o)} & A_{1,2}^{(e \backslash o)} & \cdots & A_{1,N}^{(e \backslash o)}  \\
  A_{2,0}^{(e \backslash o)} & A_{2,1}^{(e \backslash o)} & A_{2,2}^{(e \backslash o)} & \cdots & A_{2,N}^{(e \backslash o)}  \\
	\vdots & \vdots & \vdots & \ddots & \vdots \\
  A_{N-1,0}^{(e \backslash o)} & A_{N-1,1}^{(e \backslash o)} & A_{N-1,2}^{(e \backslash o)} & \cdots & A_{N-1,N}^{(e \backslash o)}  \\
\end{array}
\right| \equiv |\mathsf{\tilde{\Phi}}^{(e\backslash o)}|
\end{equation}
as the index $k_{0}$ changes. We define $\mathsf{\tilde{\Phi}}^{(e\backslash o)}$ to be the matrix we are taking the determinant of. We expect $\phi_{k_{0}}$ to at least approach $0$ so that $\phi$ is indeed in the Hilbert space $\mathcal{H} = L^{2}(\mathbb{R})$. This is a necessary but insufficient condition.

We check the behavior of $\phi_{k_{0}}$ for $k_{0}>>1$ (both for $k_{0}^{(e)}$ and $k_{0}^{(o)}$). We can restrict our attention to determinant of the matrix $\mathsf{\tilde{\Phi}}^{(e\backslash o)}$ since $\mathsf{C}_{0,0}[\mathsf{\Psi}^{(e \backslash o)}]$ is just a non-zero constant. We first assume that $|\mathsf{\tilde{\Phi}}^{(e\backslash o)}|$ is non-vanishing. If we use the cofactor expansion along the first column, we can see explicitly how $|\mathsf{\tilde{\Phi}}^{(e\backslash o)}|$ changes as the index $k_{0}$ becomes very large.
\begin{eqnarray}\label{asymk0}
&& |\mathsf{\tilde{\Phi}}^{(e\backslash o)}| = A_{0,0}^{(e \backslash o)}\mathsf{C}_{1,0}[\mathsf{\tilde{\Phi}}^{(e\backslash o)}] + A_{1,0}^{(e \backslash o)}\mathsf{C}_{2,0}[\mathsf{\tilde{\Phi}}^{(e\backslash o)}] + ... + A_{N-1,0}^{(e \backslash o)}\mathsf{C}_{N,0}[\mathsf{\tilde{\Phi}}^{(e\backslash o)}] \nonumber\\
&&\;\;\;\; \sim k_{0}^{\mp 1/4}\mathsf{C}_{1,0}[\mathsf{\tilde{\Phi}}^{(e\backslash o)}] + k_{0}^{1 \mp 1/4}\mathsf{C}_{2,0}[\mathsf{\tilde{\Phi}}^{(e\backslash o)}] + ... +  k_{0}^{N-1 \mp 1/4}\mathsf{C}_{N,0}[\mathsf{\tilde{\Phi}}^{(e\backslash o)}] \nonumber \\
\end{eqnarray}
Note that the $\mathsf{C}_{i,0}[\mathsf{\tilde{\Phi}}^{(e\backslash o)}]$'s are just constants since the first column is always removed so that they do not depend on $k_{0}$. Please refer to the appendix to show $A_{j,0}^{(e \backslash o)} \sim k_{0}^{j \mp 1/4}$ where the negative sign of $\mp$ is for $A_{j,0}^{(e)}$ and the positive sign of $\mp$ is for $A_{j,0}^{(o)}$. We see that $|\mathsf{\tilde{\Phi}}^{(e\backslash o)}|$ behaves like a polynomial in $k_{0}$ with an overall factor $k_{0}^{\mp 1/4}$ for very large $k_{0}$. This suggests that the absolute value of $|\mathsf{\tilde{\Phi}}^{(e\backslash o)}|$ becomes arbitrarily large as $k_{0}$ arbitrarily increases. Referring to equation (\ref{phirel}), if $|\mathsf{\tilde{\Phi}}^{(e\backslash o)}|$ is non-vanishing, $\phi_{k_{0}}^{(e \backslash o)*}$ becomes very large as $k_{0}>>1$ so that $\phi$ is no longer in the Hilbert Space $ L^2(\mathbb{R})$. For the free case $N=1$, we only have the overall factor $k_{0}^{\mp 1/4}$ which still leads to $\phi$ being outside $L^2(\mathbb{R})$.

We must then have $|\mathsf{\tilde{\Phi}}^{(e\backslash o)}| = 0$ at least for some maximum value of $k_{0}$ and higher. Referring back to Eqs (\ref{asymk0}),
\begin{eqnarray}\label{polyk0ev}
&& |\mathsf{\tilde{\Phi}}^{(e)}| = 0 = A_{0,0}^{(e)}\mathsf{C}_{1,0}[\mathsf{\tilde{\Phi}}^{(e)}] + A_{1,0}^{(e)}\mathsf{C}_{2,0}[\mathsf{\tilde{\Phi}}^{(e)}] + ... + A_{N-1,0}^{(e)}\mathsf{C}_{N,0}[\mathsf{\tilde{\Phi}}^{(e)}] \nonumber\\
&& 0 = f_{0,k_{0}}\sqrt{2}\sqrt[4]{\pi}\frac{\sqrt{(2k_{0})!}}{2^{k_{0}}k_{0}!}\mathsf{C}_{1,0}[\mathsf{\tilde{\Phi}}^{(e)}] + ... + f_{N-1,k_{0}}\sqrt{2}\sqrt[4]{\pi}\frac{\sqrt{(2k_{0})!}}{2^{k_{0}}k_{0}!}\mathsf{C}_{N,0}[\mathsf{\tilde{\Phi}}^{(e)}]\nonumber\\
&& 0 = f_{0,k_{0}}\mathsf{C}_{1,0}[\mathsf{\tilde{\Phi}}^{(e)}] + ... + f_{N-1,k_{0}}\mathsf{C}_{N,0}[\mathsf{\tilde{\Phi}}^{(e)}]\nonumber\\
\end{eqnarray}
\begin{eqnarray}\label{polyk0od}
&& |\mathsf{\tilde{\Phi}}^{(o)}| = 0 = A_{0,0}^{(o)}\mathsf{C}_{1,0}[\mathsf{\tilde{\Phi}}^{(o)}] + A_{1,0}^{(o)}\mathsf{C}_{2,0}[\mathsf{\tilde{\Phi}}^{(o)}] + ... + A_{N-1,0}^{(o)}\mathsf{C}_{N,0}[\mathsf{\tilde{\Phi}}^{(o)}] \nonumber\\
&& 0 = \left(f_{0,k_{0}+1} + f_{0,k_{0}}\right)\sqrt[4]{\pi}\frac{\sqrt{(2k_{0}+1)!}}{2^{k_{0}}k_{0}!}\mathsf{C}_{1,0}[\mathsf{\tilde{\Phi}}^{(o)}] + ... \nonumber\\
&& \;\;\;\;\;\;\;\;\;\;\;\;\;\;\;\;\;\;\; + \left(f_{N-1,k_{0}+1} + f_{N-1,k_{0}}\right)\sqrt[4]{\pi}\frac{\sqrt{(2k_{0}+1)!}}{2^{k_{0}}k_{0}!}\mathsf{C}_{N,0}[\mathsf{\tilde{\Phi}}^{(o)}]\nonumber\\
&& 0 = \left(f_{0,k_{0}+1} + f_{0,k_{0}}\right)\mathsf{C}_{1,0}[\mathsf{\tilde{\Phi}}^{(o)}] +  ... + \left(f_{N-1,k_{0}+1} + f_{N-1,k_{0}}\right)\mathsf{C}_{N,0}[\mathsf{\tilde{\Phi}}^{(o)}]\nonumber\\
\end{eqnarray}
where, $f_{j,k_{0}}$ satisfies the recurrence relation $f_{j,k_{0}} = (2k_{0}+1)/2\;f_{j-1,k_{0}+1} + (2k_{0}+1/2)f_{j-1,k_{0}} + k_{0}f_{j-1,k_{0}-1}$ and $f_{0,k_{0}} = 1$. Please refer to the appendix for its derivation. From the recurrence relation, we can deduce that $f_{j,k_{0}}$ is just a polynomial in $k_{0}$ of order $j$. Since the $\mathsf{C}_{j,0}[\mathsf{\tilde{\Phi}}^{(e \backslash o)}]$ are fixed constants for some fixed $\{k_{1}, k_{2}, ..., k_{N}\}$, Eqs (\ref{polyk0ev}) and (\ref{polyk0od}) can be interpreted as the vanishing of a polynomial in $k_{0}$ of order $N-1$. The number of roots of these polynomials would then be $N-1$. That is, the number of $k_{0}$'s that would make $\mathsf{\tilde{\Phi}}^{(e\backslash o)}$, and in turn, $\phi_{k_{0}}^{(e \backslash o)*}$ vanish is at most $N-1$. This is not, however, enough for $\sum_{k_{0}}|\phi_{k_{0}}^{(e \backslash o)}|^{2}$ to terminate. Therefore, $\phi$ is still outside the Hilbert Space.

Now, since the $f_{j,k_{0}}$'s, and also the $(f_{j,k_{0}+1} + f_{j,k_{0}})$, are linearly independent (that is, $\alpha_{0}f_{0,k_{0}}+\alpha_{1}f_{1,k_{0}}+...+\alpha_{N-1}f_{N-1,k_{0}}=0$ for all $k_{0}$ implies that $\alpha_{0} = \alpha_{1} = ... = \alpha_{N-1} = 0$), its coefficients, the $\mathsf{C}_{j,0}[\mathsf{\tilde{\Phi}}^{(e \backslash o)}]$'s, should vanish for all $j=1,2,3,...,N$. 
So now, we ask: Is there a fixed combination of non-vanishing $\phi_{k_{i}}^{(e \backslash o)*}$'s ($i=1,2,3,...,N$) so that all of the $\mathsf{C}_{j,0}[\mathsf{\tilde{\Phi}}^{(e \backslash o)}]$'s ($j=1,2,3,...,N$) are vanishing? Specifically, can we find a set of non-zero $\phi_{k_{i}}^{(e \backslash o)*}$'s that satisfies
\begin{eqnarray}\label{detvanish2}
&& \mathsf{C}_{j,0}[\mathsf{\tilde{\Phi}}^{(e \backslash o)}] = 0 = \left|
\begin{array}[pos]{cccc}
  \phi_{k_{1}}^{(e \backslash o)*} & \phi_{k_{2}}^{(e \backslash o)*} & \cdots & \phi_{k_{N}}^{(e \backslash o)*} \\
  A_{0,1}^{(e \backslash o)} & A_{0,2}^{(e \backslash o)} & \cdots & A_{0,N}^{(e \backslash o)}  \\
  A_{1,1}^{(e \backslash o)} & A_{1,2}^{(e \backslash o)} & \cdots & A_{1,N}^{(e \backslash o)}  \\
  \vdots & \vdots & \ddots & \vdots \\
  A_{j-2,1}^{(e \backslash o)} & A_{j-2,2}^{(e \backslash o)} & \cdots & A_{j-2,N}^{(e \backslash o)}  \\
  A_{j,1}^{(e \backslash o)} & A_{j,2}^{(e \backslash o)} & \cdots & A_{j,N}^{(e \backslash o)}  \\
  A_{j+1,1}^{(e \backslash o)} & A_{j+1,2}^{(e \backslash o)} & \cdots & A_{j+1,N}^{(e \backslash o)}  \\
	\vdots & \vdots & \ddots & \vdots \\
  A_{N-1,1}^{(e \backslash o)} & A_{N-1,2}^{(e \backslash o)} & \cdots & A_{N-1,N}^{(e \backslash o)}  \\
\end{array}
\right| \equiv |\mathsf{\tilde{\Phi}}^{(e\backslash o)}_{j}|\nonumber\\
\end{eqnarray}
(neglecting the signs) for all $j=1,2,...,N$? Where, $\mathsf{\tilde{\Phi}}^{(e\backslash o)}_{j}$ is an $N \times N$ matrix derived from $\mathsf{\tilde{\Phi}}^{(e\backslash o)}$ by deleting its first column and the $(j+1)^{th}$ row. The vanishing of this determinant suggests that the first row, the $\phi_{k_{i}}^{(e \backslash o)*}$'s, is a linear combination of the other rows. However, there is no unique linear combination of the other rows that can satisfy equation (\ref{detvanish2}) for all $j=1,2,...,N$. This is because, assuming a linear combination of the rows for a particular $j'$ was made so that $|\mathsf{\tilde{\Phi}}^{(e\backslash o)}_{j'}| = 0$, there is always at least one row that will be missing from $\mathsf{\tilde{\Phi}}^{(e\backslash o)}_{j'' \neq j'}$ that is present in the linear combination of rows satisfying $|\mathsf{\tilde{\Phi}}^{(e\backslash o)}_{j'}| = 0$, making the determinant $|\mathsf{\tilde{\Phi}}^{(e\backslash o)}_{j'' \neq j'}| \neq 0$.
The only way to make $\phi_{k_{0}}^{(e \backslash o)*} \propto |\mathsf{\tilde{\Phi}}^{(e\backslash o)}| = 0$ for at least from some maximum $k_{0}$ and above, that is, to make $\phi$ an element in the Hilbert space, is to make all the (fixed) coefficients vanish $\phi_{k_{i}}^{(e \backslash o)*} = 0$ and in turn $\phi_{k_{0}}^{(e \backslash o)*}$ for all values of $k_{0}$. We then have $\phi_{k} = 0$ for all $k=0,1,2,...$ so that $\phi = 0$. That is, for a fixed set of indices $\{k_{1}, k_{2}, ..., k_{N}\}$ the only vector orthogonal to all of the $\varphi^{(e \backslash o)}$'s (for the allowed $k_{0}$'s) is the zero vector. Also, since the fixed set of (allowed) indices are arbitrary, we would arrive at the same conclusion so that we have a complete set for any $\{k_{1}, k_{2}, ..., k_{N}\}$ ($k_{i} \neq k_{j \neq i}$). Then the set $S(\opr{T}_{-(m+1),n})$ is (over) complete.

We can now assign the linear span of $S(\opr{T}_{-(m+1),n})$ to be the domain of $\opr{T}_{-(m+1),n}$, call it $\mathcal{D}(\opr{T}_{-(m+1),n})$. Since $S(\opr{T}_{-(m+1),n})$ is a complete set, its linear span $\mathcal{D}(\opr{T}_{-(m+1),n})$ is necessarily dense in $L^2(\mathbb{R})$ so that $\opr{T}_{-(m+1),n}$ is a densely defined operator in $L^2(\mathbb{R})$.

Note that for $\opr{T}_{-(m'+1),n'}$'s of lower order, $m'+n'<N=m+n$, the conditions that the vectors $\bar{\varphi} \in \mathcal{D}(\opr{T}_{-(m'+1),n'})$ must satisfy is $\int_{-\infty}^{\infty}q^{\ell}\bar{\varphi}(q)dq=0$, where $\ell=0,1,2,..., n'+m'<N$. This means that $\mathcal{D}(\opr{T}_{-(m+1),n})$ is just a more restricted version of $\mathcal{D}(\opr{T}_{-(m'+1),n'})$, that is, $\mathcal{D}(\opr{T}_{-(m+1),n}) \subset \mathcal{D}(\opr{T}_{-(m'+1),n'})$.
In other words, for vectors $\varphi \in \mathcal{D}(\opr{T}_{-(m+1),n})$, they are then also elements of the domains of $\opr{T}_{-(m'+1),n'}$'s of lower order $\varphi \in \mathcal{D}(\opr{T}_{-(m'+1),n'})$. Linear sums of the corresponding operators are then well defined so that the domain of the sum would be the intersection of the domains of the individual terms which would be just the domain of the highest order term in the sum. 

One can also look at the case for the $\opr{T}_{-(m'+1),0}$'s, which are not part of the original discussion, we can still find a domain for them in the same way as outlined above. However, for $\opr{T}_{-1,0}$, which only requires $\int_{-\infty}^{\infty}\bar{\varphi}(q)dq=0$ for $\bar{\varphi}$ to be in its domain, we cannot strictly follow the steps above since we cannot write the vectors like Eqs (\ref{basicv}). But since $\opr{T}_{-1,0}$ is of lower order, the basic vectors of higher orders $\opr{T}_{-(m+1),n}$'s, say the free case $\opr{T}_{-1,1}$, would already be in $\mathcal{D}(\opr{T}_{-1,0})$. We can then assign the linear span of these vectors to be $\mathcal{D}(\opr{T}_{-1,0})$.

\section{Conclusion}\label{conc}
We have shown that the Bender-Dunne operators $\opr{T}_{-m,n}$, for positive integers $n$ and $m$,  are densely defined integral operators in the Hilbert space $L^2(\mathbb{R})$ by explicit construction of a dense subspace of $L^2(\mathbb{R})$ that is mapped by $\opr{T}_{-m,n}$ into $L^2(\mathbb{R})$. 
But this is just the first step in our ultimate goal of establishing the Hilbert space properties of Bender and Dunne minimal solutions, and our time of arrival operator $\opr{T}$. What we have established here is certainly required in establishing operators, such as the harmonic oscillator solution
\begin{equation}\label{ho}
\opr{T}=-\sum_{k=0}^{\infty} (-1)^k \frac{\mu^{2k+1} \omega^{2k}}{(2k+1)} \opr{T}_{-2k-1,2k+1},
\end{equation}
as legitimate Hilbert space operators in the entire configuration space. This is Bender and Dunne minimal solution and our time of arrival operator for the harmonic oscillator. Having established the $\opr{T}_{-m,n}$'s to be Hilbert space operators with domains satisfying  $D(\opr{T}_{-m,n}) \subset D(\opr{T}_{-m',n'})$ for  $(m'+n')<(m+n)$, one can clearly see that the partial sum of the right hand side of \ref{ho} is a densely defined operator. However, that is not enough to establish that the infinite series has a meaningful limit. This property is shared by all time of arrival operators for continous potential and the Bender-Dunne minimal solution. We will tackle this problem elsewhere.

\appendix
\section{Cofactor Notation}
Let $\mathsf{A}$ be an $(N+1) \times (N+1)$ given by
\begin{displaymath}
\mathsf{A} = 
\left(
\begin{array}[pos]{ccccc}
	A_{0,0} & A_{0,1} & A_{0,2} & \cdots & A_{0,N} \\
	A_{1,0} & A_{1,1} & A_{1,2} & \cdots & A_{1,N} \\
	A_{2,0} & A_{2,1} & A_{2,2} & \cdots & A_{2,N} \\
	\vdots & \vdots & \vdots & \ddots & \vdots \\
	A_{N,0} & A_{N,1} & A_{N,2} & \cdots & A_{N,N} \\
\end{array}
\right)
\end{displaymath}
A cofactor, denoted by $\mathsf{C}_{i,j}[\mathsf{A}]$, is defined here to be $(-1)^{i+j}$ times the determinant of the matrix obtained by eliminating the $(i+1)$th row and the $(j+1)$th column of the matix $\mathsf{A}$. Specifically,
\begin{displaymath}
\mathsf{C}_{0,0}[\mathsf{A}] = 
\left|
\begin{array}[pos]{ccccc}
	A_{1,1} & A_{1,2} & A_{1,3} & \cdots & A_{1,N} \\
	A_{2,1} & A_{2,2} & A_{2,3} & \cdots & A_{2,N} \\
	\vdots & \vdots & \vdots & \ddots & \vdots \\
	A_{N,1} & A_{N,2} & A_{N,3} & \cdots & A_{N,N} \\
\end{array}
\right|
\end{displaymath}
\begin{displaymath}
\mathsf{C}_{0,1}[\mathsf{A}] = 
(-1)\left|
\begin{array}[pos]{ccccc}
	A_{1,0} & A_{1,2} & A_{1,3} & \cdots & A_{1,N} \\
	A_{2,0} & A_{2,2} & A_{2,3} & \cdots & A_{2,N} \\
	\vdots & \vdots & \vdots & \ddots & \vdots \\
	A_{N,0} & A_{N,2} & A_{N,3} & \cdots & A_{N,N} \\
\end{array}
\right|
\end{displaymath}
We can infer that $\mathsf{C}_{0,i}[\mathsf{A}]$ do not depend on the first row of $\mathsf{A}$.

\section{Behavior of the $A_{j,i}^{(e \backslash o)}$'s}
For convenience, we first let $A_{j,i}^{(e \backslash o)} = \tilde{A}_{j,k_{i}}^{(e \backslash o)}$ to explicitly show the dependence on $k_{i}$. Here, the $k_{i}$'s are just labels. They do not necessarily refer to the $i^{th}$ element from the set of integers $\{k_{i}^{(e)}\}$ or $\{k_{i}^{(o)}\}$. From \cite{tabl}, we have
\begin{equation}\label{recurH}
q H_{m}(q) = \frac{1}{2}H_{m+1}(q) + m H_{m-1}(q)
\end{equation}
Let us obtain a recurrence relation for
\begin{eqnarray}
&& \tilde{A}_{j,k_{i}}^{(e)} = A_{j,i}^{(e)} = \int_{-\infty}^{\infty} q^{2j}\psi_{2k_{i}}(q)dq = (2^{2k_{i}}(2k_{i})!\sqrt{\pi})^{-1/2} \nonumber\\
&& \;\;\;\;\;\;\;\;\;\;\;\;\;\;\;\;\;\;\;\;\;\;\; \times\int_{-\infty}^{\infty} q^{2j-2}q^{2}H_{2k_{i}}(q)\exp\left(-q^{2}/2\right)dq \nonumber\\
\nonumber
\end{eqnarray}
by using equation (\ref{recurH}). We first focus on $q^{2}H_{2k_{i}}(q)$:
\begin{eqnarray}
&& q\left(q H_{2k_{i}}(q)\right) = q \left( \frac{1}{2}H_{2k_{i}+1}(q) + 2k_{i} H_{2k_{i}-1}(q) \right)\nonumber\\
&& \;\;\;\;\;\;\;\;\; = \frac{1}{2} \left(q H_{2k_{i}+1}(q) \right) + 2k_{i} \left(q H_{2k_{i}-1}(q) \right) \nonumber\\
&& \;\;\;\;\;\;\;\;\; = \frac{1}{2} \left(\frac{1}{2}H_{2k_{i}+2}(q) + (2k_{i}+1) H_{2k_{i}}(q) \right) \nonumber\\
&& \;\;\;\;\;\;\;\;\;\;\;\;\;\;\;\;\;\; + 2k_{i} \left(\frac{1}{2}H_{2k_{i}}(q) + (2k_{i}-1) H_{2k_{i}-2}(q) \right)\nonumber\\
&& \;\;\;\;\;\;\;\;\; = \frac{1}{4}H_{2k_{i}+2}(q) + (k_{i}+\frac{1}{2}+k_{i}) H_{2k_{i}}(q)+ 2k_{i}(2k_{i}-1) H_{2k_{i}-2}(q)\nonumber\\
\nonumber
\end{eqnarray}
We then have: 
\begin{eqnarray}\label{recurAe}
&& \tilde{A}_{j,k_{i}}^{(e)} = \frac{1}{2^{k_{i}}\sqrt{(2k_{i})!}\sqrt[4]{\pi}} \int_{-\infty}^{\infty}q^{2j-2} \exp\left(-q^{2}/2\right) \nonumber\\
&& \;\;\;\;\;\;\;\;\;\;\;\;\;\;\;\;\;\; \times \left( \frac{1}{4}H_{2k_{i}+2}(q) + (2k_{i}+\frac{1}{2}) H_{2k_{i}}(q)+ 2k_{i}(2k_{i}-1) H_{2k_{i}-2}(q) \right) dq \nonumber\\
&& \;\;\;\;\;\;\;\;\; = \int_{-\infty}^{\infty} \frac{1}{4}\frac{2\sqrt{(2k_{i}+2)(2k_{i}+1)}}{2^{k_{i}+1}\sqrt{(2k_{i}+2)!}\sqrt[4]{\pi}} q^{2j-2} H_{2k_{i}+2}(q) \exp\left(-q^{2}/2\right)dq  \nonumber\\
&& \;\;\;\;\;\;\;\;\;\;\;\;\;\;\;\;\;\; + \int_{-\infty}^{\infty} (2k_{i}+\frac{1}{2}) \frac{1}{2^{k_{i}}\sqrt{(2k_{i})!}\sqrt[4]{\pi}} q^{2j-2} H_{2k_{i}}(q) \exp\left(-q^{2}/2\right)dq \nonumber\\
&& \;\;\;\;\;\;\;\;\;\;\;\;\;\;\;\;\;\; + \int_{-\infty}^{\infty} \frac{2k_{i}(2k_{i}-1)2^{-1}}{\sqrt{(2k_{i})(2k_{i}-1)}2^{k_{i}-1}\sqrt{(2k_{i}-2)!}\sqrt[4]{\pi}} q^{2j-2} H_{2k_{i}-2}(q) \exp\left(-q^{2}/2\right)dq \nonumber\\
&& \;\;\;\;\;\;\;\;\; = \frac{\sqrt{(2k_{i}+2)(2k_{i}+1)}}{2}\tilde{A}_{j-1,k_{i}+1}^{(e)} + \left(2k_{i}+\frac{1}{2}\right)\tilde{A}_{j-1,k_{i}}^{(e)} +  \frac{\sqrt{2k_{i}(2k_{i}-1)}}{2}\tilde{A}_{j-1,k_{i}-1}^{(e)} \nonumber\\
\end{eqnarray}

We can do a similar procedure to a recurrence relation for 
\begin{eqnarray}
&& \tilde{A}_{j,k_{i}}^{(o)} = A_{j,i}^{(o)} = \int_{-\infty}^{\infty} q^{2j+1}\psi_{2k_{i}+1}(q)dq= (2^{2k_{i}+1}(2k_{i}+1)!\sqrt{\pi})^{-1/2} \nonumber\\
&& \;\;\;\;\;\;\;\;\;\;\;\;\;\;\;\;\;\;\;\;\;\;\;\times \int_{-\infty}^{\infty} q^{2j}(qH_{2k_{i}+1}(q))\exp\left(-q^{2}/2\right)dq\nonumber\\
\nonumber
\end{eqnarray}
in terms of $\tilde{A}_{j,k_{i}}^{(e)}$.
We have:
\begin{eqnarray}\label{recurAo}
&& \tilde{A}_{j,k_{i}}^{(o)} = \frac{1}{\sqrt{2^{2k_{i}+1}(2k_{i}+1)!\sqrt{\pi}}}\left(\frac{1}{2}\int_{-\infty}^{\infty} q^{2j}H_{2k_{i}+2}(q)\exp\left(-q^{2}/2\right)dq\right.  \nonumber\\
&& \;\;\;\;\;\;\;\;\;\;\;\;\;\;\;\;\;\;\left. + (2k_{i}+1)\int_{-\infty}^{\infty} q^{2j}H_{2k_{i}}(q)\exp\left(-q^{2}/2\right)dq \right) \nonumber\\
&& \;\;\;\;\;\;\;\;\; = \frac{1}{2}\int_{-\infty}^{\infty}\frac{\sqrt{2(2k_{i}+2)}}{\sqrt{2^{2k_{i}+2}(2k_{i}+2)!\sqrt{\pi}}} q^{2j}H_{2k_{i}+2}(q)\exp\left(-q^{2}/2\right)dq  \nonumber\\
&& \;\;\;\;\;\;\;\;\;\;\;\;\;\;\;\;\;\; + \frac{2k_{i}+1}{\sqrt{2(2k_{i}+1)}} \int_{-\infty}^{\infty}\frac{1}{\sqrt{2^{2k_{i}}(2k_{i})!\sqrt{\pi}}} q^{2j}H_{2k_{i}}(q)\exp\left(-q^{2}/2\right)dq  \nonumber\\
&& \;\;\;\;\;\;\;\;\; = \frac{\sqrt{2k_{i}+2}}{\sqrt{2}}\tilde{A}_{j,k_{i}+1}^{(e)} + \frac{\sqrt{2k_{i}+1}}{\sqrt{2}}\tilde{A}_{j,k_{i}}^{(e)} \nonumber\\
\end{eqnarray}
The recurrence relations suggests that we can obtain the $\tilde{A}_{j,k_{i}}^{(e\backslash o)}$'s by first calculating $\tilde{A}_{0,k_{i}}^{(e)} = A_{0,i}^{(e)} = (2^{k_{i}}(2k_{i})!\sqrt[4]{\pi})^{-1}\int_{-\infty}^{\infty}H_{2k_{i}}(q)\exp\left(-q^{2}/2\right)dq$. From \cite{tabl},
\begin{displaymath}
\int_{-\infty}^{\infty} \mbox{e}^{-\left(\frac{q}{\sqrt{2}}\right)^{2}} H_{2m}\left(\sqrt{2}\frac{q}{\sqrt{2}}\right) dq = \sqrt{2}\sqrt{\pi}\frac{(2m)!}{m!}
\end{displaymath}
then
\begin{equation}\label{A0}
\tilde{A}_{0,k_{i}}^{(e)} = \frac{1}{2^{k_{i}}(2k_{i})!\sqrt[4]{\pi}}\sqrt{2\pi}\frac{(2k_{i})!}{k_{i}!} = \frac{\sqrt[4]{\pi}\sqrt{(2k_{i})!}}{2^{k_{i}-1/2}k_{i}!}
\end{equation}
Using Stirling's formula $n! \sim n^{n+1/2}\exp(-n)$ for very large $n$, we get the behavior of equation (\ref{A0}) for large $k_{i}$
\begin{equation}\label{A0large}
\tilde{A}_{0,k_{i}}^{(e)} = \frac{\sqrt[4]{\pi}\sqrt{(2k_{i})!}}{2^{k_{i}-1/2}k_{i}!} \sim \frac{1}{2^{k_{i}}}\frac{(2k_{i})^{k_{i}+1/4}\exp(-k_{i})}{k_{i}^{k_{i}+1/2}\exp(-k_{i})} \sim k_{i}^{-1/4}
\end{equation}
We can use the asymptotic behavior equation (\ref{A0large}) to determine the behavior of the other $\tilde{A}_{j,k_{i}}^{(e\backslash o)}$'s. From equation (\ref{recurAe})
\begin{eqnarray}
&& \tilde{A}_{1,k_{i}}^{(e)} = \frac{\sqrt{(2k_{i}+2)(2k_{i}+1)}}{2}\tilde{A}_{0,k_{i}+1}^{(e)} + \left(2k_{i}+\frac{1}{2}\right)\tilde{A}_{0,k_{i}}^{(e)} +  \frac{\sqrt{2k_{i}(2k_{i}-1)}}{2}\tilde{A}_{0,k_{i}-1}^{(e)} \nonumber\\
&& \;\;\;\;\;\;\;\; \sim k_{i}\left(k_{i}^{-1/4}\right) + 2k_{i}\left(k_{i}^{-1/4}\right) + k_{i}\left(k_{i}^{-1/4}\right) \sim k_{i}^{3/4} \nonumber\\
&& \tilde{A}_{2,k_{i}}^{(e)} = \frac{\sqrt{(2k_{i}+2)(2k_{i}+1)}}{2}\tilde{A}_{1,k_{i}+1}^{(e)} + \left(2k_{i}+\frac{1}{2}\right)\tilde{A}_{1,k_{i}}^{(e)} +  \frac{\sqrt{2k_{i}(2k_{i}-1)}}{2}\tilde{A}_{1,k_{i}-1}^{(e)} \nonumber\\
&& \;\;\;\;\;\;\;\; \sim k_{i}\left(k_{i}^{3/4}\right) + 2k_{i}\left(k_{i}^{3/4}\right) + k_{i}\left(k_{i}^{3/4}\right) \sim k_{i}^{7/4} \nonumber\\
&& \tilde{A}_{3,k_{i}}^{(e)} = \frac{\sqrt{(2k_{i}+2)(2k_{i}+1)}}{2}\tilde{A}_{2,k_{i}+1}^{(e)} + \left(2k_{i}+\frac{1}{2}\right)\tilde{A}_{2,k_{i}}^{(e)} +  \frac{\sqrt{2k_{i}(2k_{i}-1)}}{2}\tilde{A}_{2,k_{i}-1}^{(e)} \nonumber\\
&& \;\;\;\;\;\;\;\; \sim k_{i}\left(k_{i}^{7/4}\right) + 2k_{i}\left(k_{i}^{7/4}\right) + k_{i}\left(k_{i}^{7/4}\right) \sim k_{i}^{11/4} \nonumber\\
\nonumber
\end{eqnarray}
We can infer that $\tilde{A}_{j,k_{i}}^{(e)} \sim k_{i}^{j-1/4} $.

Similarly, from equation (\ref{recurAo}) we get 
\begin{eqnarray}
&& \tilde{A}_{0,k_{i}}^{(o)} = \frac{\sqrt{2k_{i}+2}}{2}\tilde{A}_{0,k_{i}+1}^{(e)} + \frac{\sqrt{2k_{i}+1}}{2}\tilde{A}_{0,k_{i}}^{(e)} \nonumber\\
&& \;\;\;\;\;\;\;\; \sim \sqrt{k_{i}}\left(k_{i}^{-1/4}\right) + \sqrt{k_{i}}\left(k_{i}^{-1/4}\right) \sim k_{i}^{1/4}\nonumber\\
&& \tilde{A}_{1,k_{i}}^{(o)} = \frac{\sqrt{2k_{i}+2}}{2}\tilde{A}_{1,k_{i}+1}^{(e)} + \frac{\sqrt{2k_{i}+1}}{2}\tilde{A}_{1,k_{i}}^{(e)} \nonumber\\
&& \;\;\;\;\;\;\;\; \sim \sqrt{k_{i}}\left(k_{i}^{3/4}\right) + \sqrt{k_{i}}\left(k_{i}^{3/4}\right) \sim k_{i}^{5/4}\nonumber\\
&& \tilde{A}_{2,k_{i}}^{(o)} = \frac{\sqrt{2k_{i}+2}}{2}\tilde{A}_{2,k_{i}+1}^{(e)} + \frac{\sqrt{2k_{i}+1}}{2}\tilde{A}_{2,k_{i}}^{(e)} \nonumber\\
&& \;\;\;\;\;\;\;\; \sim \sqrt{k_{i}}\left(k_{i}^{7/4}\right) + \sqrt{k_{i}}\left(k_{i}^{7/4}\right) \sim k_{i}^{9/4}\nonumber\\
&& \tilde{A}_{3,k_{i}}^{(o)} = \frac{\sqrt{2k_{i}+2}}{2}\tilde{A}_{3,k_{i}+1}^{(e)} + \frac{\sqrt{2k_{i}+1}}{2}\tilde{A}_{3,k_{i}}^{(e)} \nonumber\\
&& \;\;\;\;\;\;\;\; \sim \sqrt{k_{i}}\left(k_{i}^{11/4}\right) + \sqrt{k_{i}}\left(k_{i}^{11/4}\right) \sim k_{i}^{13/4}\nonumber\\
\nonumber
\end{eqnarray}
We can then infer that $\tilde{A}_{j,k_{i}}^{(o)} \sim k_{i}^{j+1/4} $. Or more compactly, $A_{j,i}^{(e \backslash o)} = \tilde{A}_{j,k_{i}}^{(e \backslash o)} \sim k_{i}^{j\mp 1/4}$ for very large $k_{i}$.

\section{Deriving the $f_{j,k_{i}}$'s}
Let $\tilde{A}_{j,k_{i}}^{(e)} = f_{j,k_{i}} \sqrt{2}\sqrt[4]{\pi}\frac{\sqrt{(2k_{i})!}}{2^{k_{i}}k_{i}!} $. Substitute this back to equation (\ref{recurAe}) to obtain a recurrence relation for $f_{j,k_{i}}$.
\begin{eqnarray}
&& \tilde{A}_{j,k_{i}}^{(e)} = \frac{\sqrt{(2k_{i}+2)(2k_{i}+1)}}{2}\tilde{A}_{j-1,k_{i}+1}^{(e)} + \left(2k_{i}+\frac{1}{2}\right)\tilde{A}_{j-1,k_{i}}^{(e)} +  \frac{\sqrt{2k_{i}(2k_{i}-1)}}{2}\tilde{A}_{j-1,k_{i}-1}^{(e)} \nonumber\\
&& f_{j,k_{i}} \sqrt{2}\sqrt[4]{\pi}\frac{\sqrt{(2k_{i})!}}{2^{k_{i}}k_{i}!} = \frac{\sqrt{(2k_{i}+2)(2k_{i}+1)}}{2}f_{j-1,k_{i}+1} \sqrt{2}\sqrt[4]{\pi}\frac{\sqrt{(2k_{i}+2)!}}{2^{k_{i}+1}(k_{i}+1)!} \nonumber\\ 
&& \;\;\;\;\;\;\;\;\; +\left(2k_{i}+\frac{1}{2}\right)f_{j-1,k_{i}} \sqrt{2}\sqrt[4]{\pi}\frac{\sqrt{(2k_{i})!}}{2^{k_{i}}k_{i}!} +  \frac{\sqrt{2k_{i}(2k_{i}-1)}}{2}f_{j-1,k_{i}-1} \sqrt{2}\sqrt[4]{\pi}\frac{\sqrt{(2k_{i}-2)!}}{2^{k_{i}-1}(k_{i}-1)!} \nonumber\\
&& f_{j,k_{i}} \frac{\sqrt{(2k_{i})!}}{2^{k_{i}}k_{i}!} = \frac{(2k_{i}+2)(2k_{i}+1)}{4(k_{i}+1)}f_{j-1,k_{i}+1} \frac{\sqrt{(2k_{i})!}}{2^{k_{i}}k_{i}!} \nonumber\\ 
&& \;\;\;\;\;\;\;\;\; +\left(2k_{i}+\frac{1}{2}\right)f_{j-1,k_{i}} \frac{\sqrt{(2k_{i})!}}{2^{k_{i}}k_{i}!} + k_{i}f_{j-1,k_{i}-1} \frac{\sqrt{(2k_{i})!}}{2^{k_{i}}k_{i}!} \nonumber\\
&& f_{j,k_{i}} = \frac{(2k_{i}+1)}{2}f_{j-1,k_{i}+1} + \left(2k_{i}+\frac{1}{2}\right)f_{j-1,k_{i}} + k_{i}f_{j-1,k_{i}-1}  \nonumber\\ 
\nonumber
\end{eqnarray}
And from equation (\ref{A0}) we have $f_{0,k_{i}} = 1$. We can infer here that the $f_{j,k_{i}}$'s are just polynomials in $k_{i}$ of order $j$. Note that the $f_{j,k_{i}}$'s are linearly independent polynomials. That is, $\alpha_{0}f_{0,k_{i}}+\alpha_{1}f_{1,k_{i}}+\alpha_{2}f_{2,k_{i}}+...+\alpha_{N-1}f_{N-1,k_{i}}=0$ for all $k_{i}$ implies that $\alpha_{0} = \alpha_{1} = \alpha_{2} = ... = \alpha_{N-1} = 0$.

Using equation (\ref{recurAo}) we also get:
\begin{eqnarray}
&& \tilde{A}_{j,k_{i}}^{(o)} = \frac{\sqrt{2k_{i}+2}}{\sqrt{2}}\tilde{A}_{j,k_{i}+1}^{(e)} + \frac{\sqrt{2k_{i}+1}}{\sqrt{2}}\tilde{A}_{j,k_{i}}^{(e)} \nonumber\\
&& \;\;\;\;\;\;\;\;\; = \frac{\sqrt{2k_{i}+2}}{\sqrt{2}} \sqrt{2}\sqrt[4]{\pi}\frac{\sqrt{(2k_{i}+2)!}}{2^{k_{i}+1}(k_{i}+1)!}f_{j,k_{i}+1}  + \frac{\sqrt{2k_{i}+1}}{\sqrt{2}}\sqrt{2}\sqrt[4]{\pi}\frac{\sqrt{(2k_{i})!}}{2^{k_{i}}k_{i}!}f_{j,k_{i}} \nonumber\\
&& \;\;\;\;\;\;\;\;\; = \sqrt[4]{\pi}\frac{\sqrt{(2k_{i}+1)!}}{2^{k_{i}}k_{i}!}f_{j,k_{i}+1}  + \sqrt[4]{\pi}\frac{\sqrt{(2k_{i}+1)!}}{2^{k_{i}}k_{i}!}f_{j,k_{i}} \nonumber\\
&& \;\;\;\;\;\;\;\;\; = \left(f_{j,k_{i}+1} + f_{j,k_{i}} \right)\sqrt[4]{\pi}\frac{\sqrt{(2k_{i}+1)!}}{2^{k_{i}}k_{i}!} \nonumber\\
\nonumber
\end{eqnarray}
We also see that the $\left(f_{j,k_{i}+1} + f_{j,k_{i}} \right)$'s are also just polynomials in $k_{i}$ of order $j$. Since the $f_{j,k_{i}+1}$'s and $f_{j,k_{i}}$'s are of the same order, they are also linearly independent.

Explicitly, the first few $f_{j,k_{i}}$'s are
\begin{eqnarray}
&& f_{0,k_{i}} = 1 \nonumber\\
&& f_{1,k_{i}} = 4k_{i} + 1 \nonumber\\
&& f_{2,k_{i}} = 16k_{i}^{2} + 8k_{i} + 3 \nonumber\\
&& f_{3,k_{i}} = 64k_{i}^{3} + 48k_{i}^{2} + 68k_{i} + 15 \nonumber\\
&& f_{4,k_{i}} = 256k_{i}^{4} + 256k_{i}^{3} + 800k_{i}^{2} + 368k_{i} + 105 \nonumber\\
&& f_{5,k_{i}} = 1024k_{i}^{5} + 1280k_{i}^{4} + 7040k_{i}^{3} + 4960k_{i}^{2} + 4596k_{i} + 945 \nonumber\\
\nonumber
\end{eqnarray}
and for also, the $\left(f_{j,k_{i}+1} + f_{j,k_{i}} \right)$'s are
\begin{eqnarray}
&& f_{0,k_{i}+1}+f_{0,k_{i}} = 2 \nonumber\\
&& f_{1,k_{i}+1}+f_{1,k_{i}} = 8k_{i} + 6 \nonumber\\
&& f_{2,k_{i}+1}+f_{2,k_{i}} = 32k_{i}^{2} + 48k_{i} + 30 \nonumber\\
&& f_{3,k_{i}+1}+f_{3,k_{i}} = 128k_{i}^{3} + 288k_{i}^{2} + 424k_{i} + 210 \nonumber\\
&& f_{4,k_{i}+1}+f_{4,k_{i}} = 512k_{i}^{4} + 1536k_{i}^{3} + 3904k_{i}^{2} + 4128k_{i} + 1890 \nonumber\\
&& f_{5,k_{i}+1}+f_{5,k_{i}} = 2048k_{i}^{5} + 7680k_{i}^{4} + 29440k_{i}^{3} + 48960k_{i}^{2} + 50472k_{i} + 20790 \nonumber\\
\nonumber
\end{eqnarray}

\section{Explicit $|\mathsf{\tilde{\Phi}}^{(e \backslash o)}_{j}|'s$}
\begin{eqnarray}
&&  |\mathsf{\tilde{\Phi}}^{(e\backslash o)}_{1}| = \left|
\begin{array}[pos]{cccc}
  \phi_{k_{1}}^{(e \backslash o)*} & \phi_{k_{2}}^{(e \backslash o)*} & \cdots & \phi_{k_{N}}^{(e \backslash o)*} \\
  A_{1,1}^{(e \backslash o)} & A_{1,2}^{(e \backslash o)} & \cdots & A_{1,N}^{(e \backslash o)}  \\
  A_{2,1}^{(e \backslash o)} & A_{2,2}^{(e \backslash o)} & \cdots & A_{2,N}^{(e \backslash o)}  \\
  A_{3,1}^{(e \backslash o)} & A_{3,2}^{(e \backslash o)} & \cdots & A_{3,N}^{(e \backslash o)}  \\
  \vdots & \vdots & \ddots & \vdots \\
  A_{N-1,1}^{(e \backslash o)} & A_{N-1,2}^{(e \backslash o)} & \cdots & A_{N-1,N}^{(e \backslash o)}  \\
\end{array}
\right|\nonumber\\
\nonumber\\
\end{eqnarray}
\begin{eqnarray}
&&  |\mathsf{\tilde{\Phi}}^{(e\backslash o)}_{2}| = \left|
\begin{array}[pos]{cccc}
  \phi_{k_{1}}^{(e \backslash o)*} & \phi_{k_{2}}^{(e \backslash o)*} & \cdots & \phi_{k_{N}}^{(e \backslash o)*} \\
  A_{0,1}^{(e \backslash o)} & A_{0,2}^{(e \backslash o)} & \cdots & A_{0,N}^{(e \backslash o)}  \\
  A_{2,1}^{(e \backslash o)} & A_{2,2}^{(e \backslash o)} & \cdots & A_{2,N}^{(e \backslash o)}  \\
  A_{3,1}^{(e \backslash o)} & A_{3,2}^{(e \backslash o)} & \cdots & A_{3,N}^{(e \backslash o)}  \\
  A_{4,1}^{(e \backslash o)} & A_{4,2}^{(e \backslash o)} & \cdots & A_{4,N}^{(e \backslash o)}  \\
	\vdots & \vdots & \ddots & \vdots \\
  A_{N-1,1}^{(e \backslash o)} & A_{N-1,2}^{(e \backslash o)} & \cdots & A_{N-1,N}^{(e \backslash o)}  \\
\end{array}
\right|\nonumber\\
\end{eqnarray}
\begin{eqnarray}
&&  |\mathsf{\tilde{\Phi}}^{(e\backslash o)}_{3}| = \left|
\begin{array}[pos]{cccc}
  \phi_{k_{1}}^{(e \backslash o)*} & \phi_{k_{2}}^{(e \backslash o)*} & \cdots & \phi_{k_{N}}^{(e \backslash o)*} \\
  A_{0,1}^{(e \backslash o)} & A_{0,2}^{(e \backslash o)} & \cdots & A_{0,N}^{(e \backslash o)}  \\
  A_{1,1}^{(e \backslash o)} & A_{1,2}^{(e \backslash o)} & \cdots & A_{1,N}^{(e \backslash o)}  \\
  A_{3,1}^{(e \backslash o)} & A_{3,2}^{(e \backslash o)} & \cdots & A_{3,N}^{(e \backslash o)}  \\
  A_{4,1}^{(e \backslash o)} & A_{4,2}^{(e \backslash o)} & \cdots & A_{4,N}^{(e \backslash o)}  \\
  A_{5,1}^{(e \backslash o)} & A_{5,2}^{(e \backslash o)} & \cdots & A_{5,N}^{(e \backslash o)}  \\
	\vdots & \vdots & \ddots & \vdots \\
  A_{N-1,1}^{(e \backslash o)} & A_{N-1,2}^{(e \backslash o)} & \cdots & A_{N-1,N}^{(e \backslash o)}  \\
\end{array}
\right|\nonumber\\
\end{eqnarray}

\end{document}